\documentclass[conference]{IEEEtran}
\IEEEoverridecommandlockouts
\usepackage{cite}
\usepackage{amsmath,amssymb,amsfonts}
\usepackage{amsthm}
\usepackage{algorithm}
\usepackage{algcompatible}
\usepackage{graphicx}
\usepackage{bm}
\usepackage{textcomp}
\usepackage{xcolor}

\def\BibTeX{{\rm B\kern-.05em{\sc i\kern-.025em b}\kern-.08em
    T\kern-.1667em\lower.7ex\hbox{E}\kern-.125emX}}

\begin{document}
\title{Joint Activity-Delay Detection and Channel Estimation for Asynchronous Massive Random Access\\
\thanks{This work was supported by the General Research Fund (Project No. 15207220) from the Hong Kong Research Grants Council.}
}

\author{\IEEEauthorblockN{Xinyu Bian$^{\ast}$, Yuyi Mao$^{\dagger}$, and Jun Zhang$^{\ast}$}
\IEEEauthorblockA{{$^{\ast}$Dept. of ECE, The Hong Kong University of Science and Technology, Hong Kong}\\
{$^{\dagger}$Dept. of EIE, The Hong Kong Polytechnic University, Hong Kong} \\
Email: xinyu.bian@connect.ust.hk, yuyi-eie.mao@polyu.edu.hk, eejzhang@ust.hk}
}

\maketitle
\begin{abstract}
Most existing studies on joint activity detection and channel estimation for grant-free massive random access (RA) systems assume perfect synchronization among all active users, which is hard to achieve in practice. Therefore, this paper considers asynchronous grant-free massive RA systems and develops novel algorithms for joint user activity detection, synchronization delay detection, and channel estimation. In particular, the framework of orthogonal approximate message passing (OAMP) is first utilized to deal with the non-independent and identically distributed (i.i.d.) pilot matrix in asynchronous grant-free massive RA systems, and an OAMP-based algorithm capable of leveraging the common sparsity among the received pilot signals from multiple base station antennas is developed. To reduce the computational complexity, a memory AMP (MAMP)-based algorithm is further proposed that eliminates the matrix inversions in the OAMP-based algorithm. Simulation results demonstrate the effectiveness of the two proposed algorithms over the baseline methods. Besides, the MAMP-based algorithm reduces 37\% of the computations while maintaining comparable detection/estimation accuracy, compared with the OAMP-based algorithm.
\end{abstract}
  
\begin{IEEEkeywords}
Grant-free massive random access, activity detection, delay detection, channel estimation, asynchronous connectivity, approximate message passing (AMP).
\end{IEEEkeywords}

\section{Introduction}
The proliferation of Internet of Things (IoT) is pushing forward massive machine-type communications (mMTC) to provide scalable and seamless wireless connectivity \cite{cbo2016}. The most distinctive feature of mMTC is the sporadic uplink traffic pattern, i.e., only a small and random subset of users are active for transmission at each time, which entails novel random access (RA) mechanisms \cite{mt2014}. Therefore, grant-free RA, which allows users to transmit without approval from the base station (BS) \cite{ywu2020}, has been proposed as a promising solution for mMTC with low signalling overhead and access latency \cite{psch2017}.

Typically, an active user in a grant-free massive RA system directly transmits a unique pilot sequence for user activity detection and channel estimation at the BS \cite{xchen2021}. However, due to the massive potential users, only non-orthogonal pilot sequences can be adopted, which brings substantial difficulties to user activity detection and channel estimation. Fortunately, they appear to be compressive sensing (CS) problems \cite{donoho2006} because of the sporadic traffic pattern, for which many efficient algorithms are available \cite{trob1996,jat2004,donoho2010}.

In general, two kinds of methods have been developed for user activity detection and channel estimation in grant-free massive RA systems. The first kind of methods solve a maximum likelihood (ML) estimation problem based on the sample covariance matrix of the received pilot signal, which serves as a sufficient statistic for the user activity \cite{yafeng2023}. For example, the set of active users were detected via coordinate-wise descent in \cite{sha2018} and \cite{zchen2019}. However, in such methods, channel estimation needs to be performed after user activity detection and the computational complexity is usually high. On the other hand, the second kind of methods perform joint activity detection and channel estimation (JADCE) by invoking the family of approximate message passing (AMP) algorithms, which enjoys much lower complexity at the cost of some performance degradation \cite{zchen2018, mke2020, xbian2022}. Specifically, an AMP algorithm with a minimum mean square error (MMSE) denoiser was proposed for JADCE in \cite{zchen2018}, which was extended in \cite{mke2020} by incorporating the spatial and angular domain channel characteristics to improve accuracy. Besides, joint activity detection, channel estimation, and data decoding was investigated via the bilinear generalized AMP (BiG-AMP) algorithm in \cite{xbian2022}, which leverages the common sparsity in the received pilot and data signal, as well as the soft information from a channel decoder.

Nevertheless, the above studies assume all users are perfectly synchronized, which is hard to achieve in mMTC systems with many uncoordinated low-end IoT devices. This is because the pilot sequence sent by each active user may be randomly delayed by some unknown symbol periods \cite{hongyuan2008}. The critical asynchronous massive RA system has also received recent attention \cite{lliu2021, wzhu2021}. In \cite{lliu2021}, by formulating the problem of joint activity-delay detection and channel estimation as a group LASSO problem, a block coordinate descent algorithm was proposed, which has closed-form solutions for each block of variables. However, this algorithm does not leverage the common sparsity among the received pilot signals of multiple BS antennas and thus cannot unleash the system performance. In \cite{wzhu2021}, a learned AMP network was developed as a model-driven deep learning approach for asynchronous grant-free massive RA systems, where iterations of the AMP algorithm are unrolled as neural network layers. However, because of the delayed pilot symbols, entries of the effective pilot matrix in asynchronous grant-free massive RA systems are not independent and identically distributed (i.i.d.) according to a Gaussian distribution. Therefore, such an AMP-based algorithm significantly compromises the accuracy of activity-delay detection and channel estimation.

To narrow the research gaps, in this paper, we first propose a novel joint activity-delay detection and channel estimation algorithm for asynchronous grant-free massive RA systems based on orthogonal AMP (OAMP) \cite{jma2017}, which is suitable for general pilot matrix with non-i.i.d. Gaussian entries. The common sparsity among multiple BS antennas is also utilized to boost the detection/estimation performance. To reduce the computational complexity, we then develop a low-cost algorithm based on the memory AMP (MAMP) framework, where the high-complexity optimal linear estimator in OAMP is replaced with a memory alternative that recycles the intermediate information from previous iterations. Simulation results demonstrate the significant accuracy improvements in terms of activity detection, delay detection, and channel estimation achieved by the two proposed algorithms compared with the baseline methods. Besides, the MAMP-based algorithm maintains the performance of the OAMP-based algorithm and enjoys a similar complexity as the conventional AMP-based algorithm.

\textbf{Notations:} We use lower-case letters, bold-face lower-case letters, bold-face upper-case letters, and math calligraphy letters to denote scalars, vectors, matrices, and sets, respectively. The transpose and conjugate transpose of a matrix $\mathbf{M}$ are denoted as $\mathbf{M}^{\mathrm{T}}$ and $\mathbf{M}^{\mathrm{H}}$, respectively. Besides, we denote the complex Gaussian distribution with mean $\bm{\mu}$ and covariance matrix $\bm{\Sigma}$ as $\mathcal{C} \mathcal{N}(\bm{\mu}, \bm{\Sigma})$, and the probability density function of a complex Gaussian variable $\bm{x}$ as $\mathcal{C} \mathcal{N}(\bm{x};\bm{\mu}, \bm{\Sigma})$. In addition, $\delta_{0}$ denotes the Dirac delta function, “$\otimes$” stands for the Kronecker product, $\operatorname{tr}(\cdot)$ returns the trace of a matrix, and $\mathbb{E}[\cdot]$ and $\operatorname{Var}[\cdot]$ denote the statistical expectation and variance, respectively.

\section{System Model}
We consider the uplink communication procedure in grant-free massive RA system, which consists of $N$ single-antenna users and an $M$-antenna BS. The set of users is denoted as $\mathcal{N}\triangleq \{1,\cdots,N\}$, and the set of BS antennas is denoted as $\mathcal{M}\triangleq\{1,\cdots,M\}$. The quasi-static block fading channel model is adopted, in which channel coefficients remain constant within a transmission block, but varies independently from block to block. The uplink channel vector from user $n$ to the BS is denoted as $\mathbf{f}_{n}$. In this paper, we focus on the Rayleigh fading channels, i.e., $\mathbf{f}_{n}\sim \mathcal{CN}(\mathbf{0},\beta_{n}\mathbf{I}_{M})$, where $\beta_{n}$ denotes the large-scale fading coefficient for user $n$ known at the BS. Because of the sporadic traffic pattern, at each channel block, $K$ ($K\leq N$) out of the $N$ users are assumed to be active, and all the users become active with probability $\lambda$. Let $u_{n} \in \{0,1\}$ be the activity indicator for user $n$, where $u_{n}=1$ indicates that user $n$ is active and vice versa. Thus, the set of active users is given as $\mathcal{K} \triangleq \left\{n \in \mathcal{N} | u_{n}=1 \right\}$.

The classical two-phase grant-free RA scheme is adopted, which contains a pilot transmission phase with $\bar{L}$ symbols, followed by a data delivery phase. In the pilot transmission phase, each user is assigned with a unique pilot sequence $\sqrt{\bar{L}}\bar{\mathbf{p}}_{n}$, where $\bar{\mathbf{p}}_{n}\triangleq \left[\bar{p}_{n, 1}, \cdots, \bar{p}_{n, \bar{L}}\right]^{\mathrm{T}}$ and $\bar{p}_{n, l} \sim \mathcal{C} \mathcal{N}\left(0, 1/\bar{L}\right)$. It was shown that such a design of pilot sequences achieves asymptotic orthogonality when $\bar{L}\rightarrow \infty$ \cite{zchen2018}. Unlike most existing works on grant-free massive RA, we relax the perfect synchronization requirement and assume each user transmits the pilot sequence with a delay of some unknown symbol periods. We use $t_{n}$ to denote the unknown delay for user $n$, where $t_{n}$ is an integer uniformly distributed in set $\{0,\cdots,T\}$, and $T$ denotes the maximum symbol delay known at the BS. Therefore, the expanded pilot sequence of user $n$ with delay $t_{n}$, denoted as $\tilde{\mathbf{p}}_{n,t_{n}}$, can be expressed as $\tilde{\mathbf{p}}_{n,t_{n}}=[\mathbf{0}_{\tau_{n}}^{\mathrm{T}},\bar{\mathbf{p}}_{n}^{\mathrm{T}},\mathbf{0}_{T-t_{n}}^{\mathrm{T}}]^{\mathrm{T}}$, which is a sequence with length $L=\bar{L}+T$ obtained by padding $t_{n}$ and $T-t_{n}$ zeros before and after $\bar{\mathbf{p}}_{n}$, respectively. Correspondingly, we define $\mathbf{P}_{n}\triangleq [\tilde{\mathbf{p}}_{n,0},\cdots,\tilde{\mathbf{p}}_{n,T}] \in \mathbb{C}^{L\times (T+1)}$. 

Since both the user activity and synchronization delay need to be detected, we further introduce indicator $\boldsymbol{\phi}_{n}\triangleq [\phi_{n,0},\cdots,\phi_{n,T}]^{\mathrm{T}}$ for user $n$, where $\phi_{n,t}=1$ only when $u_{n}=1$ and $t=t_{n}$; Otherwise $\phi_{n,t}=0$. The received pilot signal $\tilde{\mathbf{Y}} \in \mathbb{C}^{L\times M}$ at the BS can be expressed as follows: 
\begin{align}
\tilde{\mathbf{Y}}&=\sqrt{\rho L}\mathbf{P}\mathbf{H}+\tilde{\mathbf{N}}, \label{eqsystemmodel}
\end{align}
\noindent where $\mathbf{P}\triangleq [\mathbf{P}_{1},\cdots,\mathbf{P}_{N}] \in \mathbb{C}^{L\times(T+1)N}$ concatenates the expanded pilot matrix of all users, and $\mathbf{H} \triangleq \left[\mathbf{H}_{1},...,\mathbf{H}_{N}\right]^{T} \in \mathbb{C}^{(T+1)N\times M}$ stands for the expanded effective channel matrix with $\mathbf{H}_{n}\triangleq \boldsymbol{\phi_{n}}\otimes \mathbf{f}_{n} \in \mathbb{C}^{(T+1)\times M}$. Besides, $\rho$ is the transmit power, and $\mathbf{N}=\left[\mathbf{n}_{1},...,\mathbf{n}_{L}\right]^{\mathrm{T}}$ denotes the Gaussian noise with zero mean and variance $\sigma^{2}$ for each element. We also define $\mathbf{Y}\triangleq \tilde{\mathbf{Y}} \slash \sqrt{\rho L}$ and $\mathbf{N}\triangleq \tilde{\mathbf{N}} \slash \sqrt{\rho L}$ as the normalized received signal and noise for ease of notation.

Based on the signal model in (\ref{eqsystemmodel}), our goal is to detect the user activity and synchronization delay, and estimate the effective channel coefficients, given the pilot sequences $\mathbf{P}$ at the BS. In the next section, we will develop a novel algorithm based on OAMP to achieve this goal.

\section{The Proposed OAMP-based Algorithm}
Conventionally, JADCE problems for synchronous grant-free massive RA can be solved via the AMP algorithm, which calculates the posterior distribution of the effective channel matrix $\mathbf{H}$ in an iterative manner. A prerequisite of the AMP algorithm is that entries of the pilot matrix $\mathbf{P}$ should be independent and identically Gaussian distributed \cite{donoho2010}, which, however, cannot be satisfied with asynchronicity. In other words, applying the AMP algorithm to our problem may result in inaccurate estimation of the effective channel matrix, and thus the user activity and synchronization delay. To tackle this limitation, the framework of OAMP \cite{jma2017} emerges as an ideal candidate. Nevertheless, the original OAMP framework was proposed to solve single measurement vector (SMV) problems, which fails to exploit the common sparsity among the multiple measurements of the BS antennas \cite{ymei2022}. In the following, we first develop an OAMP-based algorithm in Section \ref{SMV} that leverages the received signal of an individual BS antenna to perform joint activity-delay detection and channel estimation. The algorithm is then extended for multiple BS antennas in Section \ref{extendtoMMV}.

\subsection{OAMP-based Algorithm for Individual BS Antenna \label{SMV}}
We first derive the OAMP-based algorithm to recover the expanded effective channel vector for individual BS antenna based on its normalized received pilot signal given as follows:
\begin{align}
\mathbf{y}_{m}=\mathbf{P}\mathbf{h}_{m}+\mathbf{n}_{m}, \forall m \in \mathcal{M}, 
\label{SMVmodel}
\end{align}
\noindent where $\mathbf{y}_{m}$, $\mathbf{h}_{m}$, and $\mathbf{n}_{m}$ denote the $m$-th column of $\mathbf{Y}$, $\mathbf{H}$, and $\mathbf{N}$, respectively. The conventional OAMP algorithm iterates between a linear estimator (LE) and a non-linear estimator (NLE) under certain orthogonality constraints. Starting with the initialization $\mathbf{s}_{m}^{(1)}=\mathbf{0}$, the optimal OAMP structure in the $i$-th iteration is expressed as follows:
\begin{align}
\text{LE:} \ \ \ \ \ \mathbf{r}_{m}^{(i)}=\mathbf{s}_{m}^{(i)}+\mathbf{W}_{m}^{(i)}(\mathbf{y}_{m}-\mathbf{P}\mathbf{s}_{m}^{(i)}),\label{LE}
\end{align}
\begin{align}
\begin{aligned}
&\text{NLE:} \ \ \ \ \ \mathbf{s}_{m}^{(i+1)}=\eta^{(i)}(\mathbf{r}_{m}^{(i)})\\ &=C_{m}^{(i)}\Big(\hat{\eta}^{(i)}(\mathbf{r}_{m}^{(i)})-\frac{\sum_{n=1}^{N}\sum_{t=0}^{T}\hat{\eta}^{\prime(i)}(r_{n,t,m}^{(i)})}{(T+1)N}\mathbf{r}_{m}^{(i)}\Big),\label{NLE}
\end{aligned}
\end{align}
\noindent where $\mathbf{r}^{\left(i\right)}_{m}$ and $\mathbf{s}_{m}^{\left(i+1\right)}$ are respectively the output of the LE and NLE in the $i$-th iteration, and other notations will be introduced in the sequel.

\subsubsection{LE}
As shown in (\ref{LE}), the LE is applied to the normalized received signal $\mathbf{y}_{m}$ that decorrelates the vector estimation problem to $N$ scalar estimation problems for each user. This is achieved by restricting $\mathbf{W}_{m}^{(i)}$ as the following the optimal structure:
\begin{align}
\mathbf{W}_{m}^{(i)}=\frac{(T+1) N}{\operatorname{tr}\left(\hat{\mathbf{W}}_{m}^{(i)} \mathbf{P}\right)} \hat{\mathbf{W}}_{m}^{(i)},
\label{decor}
\end{align}
\noindent where $\hat{\mathbf{W}}_{m}^{(i)}=\mathbf{P}^{\mathrm{H}}\left(\mathbf{P} \mathbf{P}^{\mathrm{H}}+(\sigma^2/\rho L (v_{m}^{(i)})^{2}) \mathbf{I}_{L}\right)^{-1}$ is the linear minimum mean square error (LMMSE) estimator with $(v_{m}^{(i)})^{2}$ representing the error of the NLE that can be calculated as follows:
\begin{align}
(v_{m}^{(i)})^{2}\triangleq\frac{\mathbb{E}[||\mathbf{s}_{m}^{(i)}-\mathbf{h}_{m}||_{2}^{2}]}{(T+1)N}\approx \frac{|| \mathbf{y}_m-\mathbf{P} \mathbf{s}_m^{(i)}||^2-\sigma^{2}/\rho}{\operatorname{tr}\left(\mathbf{P}^\mathrm{H} \mathbf{P}\right)}.
\label{errorv}
\end{align}
\noindent Note that the approximation in (\ref{errorv}) is an empirical estimation \cite{jma2017}. Similarly, the error of the LE, which is defined as $(\tau_{m}^{(i)})^{2}\triangleq\frac{\mathbb{E}[||\mathbf{r}_{m}^{(i)}-\mathbf{h}_{m}||_{2}^{2}]}{(T+1)N}$, can be empirically estimated as follows:
\begin{align}
(\tau_{m}^{(i)})^{2}\!\approx\!\frac{\operatorname{tr}\left(\mathbf{B}_{m}^{(i)} (\mathbf{B}_{m}^{(i)})^{\mathrm{H}}\right) (v_{m}^{(i)})^2\!+\!\operatorname{tr}\left(\mathbf{W}_{m}^{(i)} (\mathbf{W}_{m}^{(i)})^{\mathrm{H}}\right) \cdot \frac{\sigma^2}{\rho L}}{(T+1) N} ,
\label{errortau}
\end{align}
\noindent where $\mathbf{B}_{m}^{(i)}\triangleq \mathbf{I}_{(T+1)N}-\mathbf{W}_{m}^{(i)}\mathbf{P}$.

\subsubsection{NLE} The NLE applies a denoiser $\eta^{(i)}(\cdot)$ to the output of the LE, which should be restricted to be divergence-free, i.e., $\mathbb{E}\left[\eta^{\prime(i)}(\cdot)\right] = 0$ with $\eta^{\prime(i)}(\cdot)$ denoting the ﬁrst-order derivative of $\eta^{(i)}(\cdot)$, as shown in (\ref{NLE}). In the OAMP framework, $\mathbf{r}_{m}^{(i)}$ is modeled as an observation of $\mathbf{h}_{m}$ with additive white Gaussian noise (AWGN), i.e., $\mathbf{r}_{m}^{(i)}=\mathbf{h}_{m}+\tau_{m}^{(i)}\mathbf{z}_{m}$, where $\mathbf{z}_{m}\sim \mathcal{CN}(\mathbf{0},\mathbf{I}_{(T+1)N})$ and is independent of $\mathbf{h}_{m}$. Therefore, $\hat{\eta}^{(i)}(\cdot)$ in the optimal NLE is the element-wise MMSE denoiser given as follows:
\begin{align}
\hat{\eta}^{(i)}(r_{n,t,m}^{(i)})&=\mathbb{E}[h_{n,t,m} \mid r_{n,t,m}^{(i)}],
\label{MMSEmean}
\end{align}
\noindent where $r_{n,t,m}^{(i)}$ and $h_{n,t,m}$ denotes the $t$-th entries of $\mathbf{r}_{n,m}^{(i)}$ and $\mathbf{h}_{n,m}$, respectively, and $\mathbf{r}_{m}^{(i)}=[(\mathbf{r}_{1,m}^{(i)})^{\mathrm{T}},\cdots,(\mathbf{r}_{N,m}^{(i)})^{\mathrm{T}}]^{\mathrm{T}}$, $\mathbf{h}_{m}=[\mathbf{h}_{1,m}^{\mathrm{T}},\cdots,\mathbf{h}_{N,m}^{\mathrm{T}}]^{\mathrm{T}}$. As shown in (\ref{MMSEmean}), the estimation result of the MMSE denoiser in the $i$-th iteration is the posterior mean of $h_{n,t,m}$ given $r_{n,t,m}^{(i)}$, and thus the posterior variance of $h_{n,t,m}$ is calculated as follows:
\begin{align}
\psi_{n,t,m}^{(i)}=\operatorname{Var}[h_{n,t,m} \mid r_{n,t,m}^{(i)}]&=\mathbb{E}[||\hat{\eta}^{(i)}(r_{n,t,m}^{(i)})-h_{n,t,m}||^{2}].
\label{MMSEvar}
\end{align}

\noindent Besides, $C_{m}^{(i)}$ in the optimal NLE is given as follows:
\begin{align}
C_{m}^{(i)}=\frac{(\tau_{m}^{(i)})^{2}}{(\tau_{m}^{(i)})^{2}-\bar{\psi}_{m}^{(i)}},
\label{Cm}
\end{align}
where $\bar{\psi}_{m}^{(i)}\triangleq \frac{\sum_{n=1}^{N}\sum_{t=0}^{T}\psi_{n,t,m}^{(i)}}{(T+1)N}$. The term $\frac{\sum_{n=1}^{N}\sum_{t=0}^{T}\hat{\eta}^{\prime(i)}(r_{n,t,m}^{(i)})}{(T+1)N}$ with $\hat{\eta}^{\prime(i)}(\cdot)$ denoting the ﬁrst-order derivative of $\hat{\eta}^{(i)}(\cdot)$ is derived as follows:
\begin{align}
\frac{\sum_{n=1}^{N}\sum_{t=0}^{T}\hat{\eta}^{\prime(i)}(r_{n,t,m}^{(i)})}{(T+1)N}=\frac{\bar{\psi}_{m}^{(i)}}{(\tau_{m}^{(i)})^{2}}.
\label{bigterm}
\end{align}

In the NLE, the keys are the calculations of the posterior mean and variance of the MMSE denoiser in (\ref{MMSEmean}) and (\ref{MMSEvar}).To obtain these values, prior information of $\mathbf{h}_{m}$ is required. In particular, since all the users become active with equal probability and the symbol delay of an active user is uniformly distributed, we model the prior information of $h_{n,t,m}$ for user $n$ as follows:
\begin{align}
\begin{aligned}
p\left(h_{n,t,m}\right)&=\left(1-\frac{\lambda}{T+1}\right) \delta_{0}\left(h_{n, t,m}\right)\\
&+\frac{\lambda}{T+1} \mathcal{C N}\left(h_{n,t,m} ; 0, \beta_{n}\right).
\label{prior}
\end{aligned}
\end{align}
\noindent With the prior information and the model of $\mathbf{r}_{m}^{(i)}$, the posterior distribution of $h_{n,t,m}$ can be calculated as follows:
\begin{align}
\begin{aligned}
p(h_{n,t,m}|r_{n,t,m}^{(i)})&=\left(1-\pi_{n,t,m}^{(i)}\right) \delta_{0}\left(h_{n, t,m}\right)\\
&+\pi_{n,t,m}^{(i)} \mathcal{C N}\left(h_{n,t,m} ; \mu_{n,t,m}^{(i)}, \Gamma_{n,t,m}^{(i)}\right),
\end{aligned}
\end{align}
\noindent where $\mu_{n,t,m}^{(i)}=\frac{\beta_{n}r_{n,t,m}^{(i)}}{(\tau_{m}^{(i)})^{2}+\beta_{n}}$, $\Gamma_{n,t,m}^{(i)}=\frac{(\tau_{m}^{(i)})^{2}\beta_{n}}{(\tau_{m}^{(i)})^{2}+\beta_{n}}$, and $\pi_{n,t,m}^{(i)}=\frac{\frac{\lambda}{T+1}}{\frac{\lambda}{T+1}+(1-\frac{\lambda}{T+1})e^{\xi_{n,t,m}^{(i)}}}$ with $\xi_{n,t,m}^{(i)}=\frac{|r_{n,t,m}^{(i)}|^{2}}{\beta_{n}+(\tau_{m}^{(i)})^{2}}-\frac{|r_{n,t,m}^{(i)}|^{2}}{(\tau_{m}^{(i)})^{2}}-\ln\frac{(\tau_{m}^{(i)})^{2}}{\beta_{n}+(\tau_{m}^{(i)})^{2}}$. Therefore, the posterior mean (\ref{MMSEmean}) and variance (\ref{MMSEvar}) are given as follows:
\begin{align}
\hat{\eta}^{(i)}(r_{n,t,m}^{(i)})=\pi_{n,t,m}^{(i)}\mu_{n,t,m}^{(i)},
\label{postmeancal}
\end{align}
\begin{align}
\psi_{n,t,m}^{(i)}=\pi_{n,t,m}^{(i)}(|\mu_{n,t,m}^{(i)}|^{2}+\Gamma_{n,t,m}^{(i)})-|\hat{\eta}^{(i)}(r_{n,t,m}^{(i)})|^{2}.
\label{postvarcal}
\end{align}

As a result, the effective channel vector is estimated as $\hat{\mathbf{h}}_{m}=\hat{\eta}^{(I)}(\mathbf{r}_{m}^{(I)})$, where $I$ denotes the iteration index when the OAMP iteration is terminated.

\subsection{Extension for Multiple BS Antennas \label{extendtoMMV}}
The OAMP-based algorithm developed in Section \ref{SMV} neglects the common sparsity among the received pilot signals of multiple BS antennas, which could be used to enhance the detection/estimation performance. Since all the BS antennas receive pilot signal at the same instant, it is reasonable to form a common activity indicator to update the prior information. Specifically, we change the constant the sparsity ratio $\frac{\lambda}{T+1}$ in (\ref{prior}) to variable $\omega_{n,t}^{\left(i\right)}$, which is updated according to the posterior sparsity ratio $\pi_{n,t,m}^{(i-1)}, \forall m$ in the $i$-th OAMP iteration by taking into account all the $M$ BS antennas as follows:
\begin{align}
\omega_{n,t}^{(i)}=\frac{1}{M}\sum_{m=1}^{M}\pi_{n,t,m}^{(i-1)}. \label{sparsityratio}
\end{align}

\noindent Thus, the updated prior information in the $i$-th iteration is given as follows:
\begin{align}
\begin{aligned}
p\left(h_{n,t,m}\right)&=\left(1-\omega_{n,t}^{(i)}\right) \delta\left(h_{n, t,m}\right)\\
&+\omega_{n,t}^{(i)} \mathcal{C N}\left(h_{n,t,m} ; 0, \beta_{n}\right).
\label{refinedprior}
\end{aligned}
\end{align}

\noindent Accordingly, we have $\pi_{n,t,m}^{(i)}= \frac{\omega_{n,t}^{(i)}}{\omega_{n,t}^{(i)}+(1-\omega_{n,t}^{(i)})e^{\xi_{n,t,m}^{(i)}}}$.

After the OAMP-based algorithm terminates at the $I$-th iteration, the estimated effective channel vector is obtained as $\hat{\mathbf{h}}_{m}=\hat{\eta}^{(I)}(\mathbf{r}_{m}^{(I)})$ according to Section \ref{SMV}, and the estimated set of active users is determined as $\hat{\mathcal{K}}\triangleq \{n\in \mathcal{N}|\sum_{t=0}^{T} \omega_{n,t}^{(I+1)}\geq \theta\}$, where $\theta$ is an empirical threshold. Besides, the synchronization delay for an estimated active user can be determined by $t_{n}=t_{n,\max}$, $n\in \mathcal{\hat{K}}$, where $t_{n,\max}\!=\!\mathop{\arg\max}\limits_{t \in \{0,\cdots,T\}} \omega_{n,t}^{(I+1)}$. Details of the proposed OAMP-based algorithm for joint activity-delay detection and channel estimation are summarized in Algorithm \ref{algo1}.
\addtolength{\topmargin}{0.04in}
\begin{algorithm}[htpb]
\caption{The Proposed OAMP-based Algorithm \label{algo1}}
{\bf Input:}
The normalized received pilot signal $\mathbf{Y}$, pilot sequences $\mathbf{P}$, maximum number of iterations $Q_{1}$, and accuracy tolerance $\epsilon_{1}$.\\
{\bf Output:}
The estimated effective channel matrix $\hat{\mathbf{H}}$, set of active users $\hat{\mathcal{K}}$, and synchronization delay $\{t_{n}\}$'s, $n \in \hat{\mathcal{K}}$.\\
{\bf Initialize:}
$i \leftarrow 0$, $\mathbf{s}_{m}^{(1)}=\mathbf{0}$, $m\in \mathcal{M}$, $\omega_{n,t}^{(1)}=\frac{\lambda}{T+1}$, $n \in \mathcal{N}$, $t \in \{0,\cdots,T\}$.
\begin{algorithmic}[1]
\WHILE{{$i < Q_{1}$} \text{and} {$\frac{\sum_{n,m}\sum_{t}|{s}_{n,t,m}^{(i)}-{s}_{n,t,m}^{(i-1)}|^{2}}{\sum_{n,m}\sum_{t}|{s}_{n,t,m}^{(i-1)}|^{2}} > \epsilon_{1}$}}
\STATE $i \leftarrow i+1$
\Statex \quad //\textit{The LE}//
\STATE Calculate $\mathbf{W}_{m}^{(i)}$, $\forall m$ according to (\ref{decor}).
\STATE Perform the linear estimation to obtain $\mathbf{r}_{m}^{\left(i\right)}, \forall m$ via (\ref{LE}).
\STATE Calculate $(v_{m}^{(i)})^{2}$ and $(\tau_{m}^{(i)})^{2}$, $\forall m$ according to (\ref{errorv}) and
\Statex \quad (\ref{errortau}), respectively.
\Statex \quad //\textit{The NLE}//
\STATE Enforcing the common sparsity to update the prior
\Statex \quad sparsity ratio $\omega_{n,t}^{(i)}$, $\forall n,t$ and the corresponding prior 
\Statex \quad information $p(h_{n,t,m})$, $\forall n,t,m$ according to (\ref{sparsityratio}) and 
\Statex \quad (\ref{refinedprior}), respectively.
\STATE Calculate the posterior mean $\hat{\eta}^{(i)}(r_{n,t,m}^{(i)})$ and variance
\Statex \quad $\psi_{n,t,m}^{(i)}$, $\forall n,t,m$ according to (\ref{postmeancal}) and (\ref{postvarcal}), respec-
\Statex \quad tively.
\STATE Calculate $C_{m}^{(i)}$ and $\frac{\sum_{n=1}^{N}\sum_{t=0}^{T}\hat{\eta}^{\prime(i)}(r_{n,t,m}^{(i)})}{(T+1)N}$, $\forall m$ 
\Statex \quad according to (\ref{Cm}) and (\ref{bigterm}), respectively.
\STATE Perform the non-linear estimation to obtain $\mathbf{s}_{m}^{(i+1)}$, $\forall m$ 
\Statex \quad according to (\ref{NLE}).
\ENDWHILE
\STATE Obtain $\hat{\mathbf{H}}=[\hat{\mathbf{h}}_{1},\cdots,\hat{\mathbf{h}}_{M}]$ with $\hat{\mathbf{h}}_{m}=\hat{\eta}^{(I)}(\mathbf{r}_{m}^{(I)})$.
\STATE Determine the set of active users as $\hat{\mathcal{K}}\triangleq \{n\in \mathcal{N}|\sum_{t=0}^{T} \omega_{n,t}^{(I+1)}\geq \theta\}$.
\STATE Determine the symbol delay as $t_{n}=\mathop{\arg\max}\limits_{t \in \{0,\cdots,T\}} \omega_{n,t}^{(I+1)}$.
\end{algorithmic}
\end{algorithm}

\section{Acceleration With Memory AMP}
Although the OAMP-based algorithm is effective in solving the joint activity-delay detection and channel estimation problem, the high-complexity LMMSE estimator limits its application in practical mMTC systems. Recently, the framework of MAMP was proposed to reduce the complexity caused by the LMMSE estimator in OAMP \cite{leiliu2022}. This framework was motivated by the iterative LMMSE algorithm for coded MIMO systems \cite{ychi2019}, where the costly message passing decoding is replaced by low-complexity memory decoding by utilizing information obtained in previous iterations. MAMP introduces a similar memory mechanism to OAMP, which is adopted to accelerate joint activity-delay detection and channel estimation in this section.

The MAMP-based algorithm also iterates between an LE and an NLE to process the normalized received pilot signal at the $m$-th BS antenna. The only difference from the OAMP-based algorithm is that the LE is replaced by a memory alternative. In particular, starting with $\mathbf{s}_{m}^{(1)}=\mathbf{0}$ and $\hat{\mathbf{r}}_{m}^{(0)}=\mathbf{0}$, the memory LE of the optimal MAMP structure is expressed as follows:
\begin{subequations}
\begin{align}
\hat{\mathbf{r}}_{m}^{(i)}=\iota^{(i)} \mathbf{D} \hat{\mathbf{r}}_{m}^{(i-1)}+\alpha^{(i)}\left(\mathbf{y}_{m}-\mathbf{P} \mathbf{s}_{m}^{(i)}\right),\label{MAMPLEA}\\
\mathbf{r}_{m}^{(i)}=\frac{1}{\varepsilon^{(i)}}\left(\mathbf{P}^{\mathrm{H}}\hat{\mathbf{r}}_{m}^{(i)}-\sum_{g=1}^{i}p_{(g)}^{(i)}\mathbf{s}_{m}^{(g)}\right).\label{MAMPLEB}
\end{align}
\end{subequations}

Specifically, $\iota^{(i)}=(\lambda^{\dag}+\sigma^2/\rho L (v_{m}^{(i)})^{2})^{-1}$ in (\ref{MAMPLEA}) is a relaxation parameter to improve the convergence speed of MAMP, and $\lambda^{\dag}$ is the average of the smallest and largest eigenvalues of $\mathbf{P}\mathbf{P}^{\mathrm{H}}$. Besides, $\mathbf{D}=\lambda^{\dag}\mathbf{I}-\mathbf{P}\mathbf{P}^{\mathrm{H}}$, and $\alpha^{(i)}$ is a weight factor that adjusts the contribution of $\mathbf{s}_{m}^{(i)}$ to the estimate $\mathbf{r}_{m}^{(i)}$. The optimal $\alpha^{(i)}$ can be obtained by minimizing the error of LE, which is given by $\alpha^{(1)}=1$ and
\begin{align}
\alpha^{(i)}= \frac{c_{2}^{(i)} c_{0}^{(i)}+c_{3}^{(i)}}{c_{1}^{(i)} c_{0}^{(i)}+c_{2}^{(i)}}, i>1. \label{optalpha}
\end{align}

\noindent To introduce the notations in (\ref{optalpha}), we first define for $i,j \geq 0$, $f_{i}\triangleq \frac{1}{(T+1)N} \operatorname{tr}\left\{\left(\mathbf{P} \mathbf{P}^{\mathrm{H}}\right)^i\right\}$, $b_{i}\triangleq \sum_{g=0}^{i} \binom{i}{g}(-1)^{g}(\lambda^{\dag})^{i-g}f_{i}$, $w_{i}\triangleq \lambda^{\dag} b_{i}-b_{i+1}$, $\bar{w}_{i, j} \triangleq \lambda^{\dagger} w_{i+j}-w_{i+j+1}-w_i w_j$. We also define, for $1\leq g\leq i$,
\begin{align}
\vartheta_{(g)}^{(i)} \triangleq \begin{cases}\alpha^{(i)}, & g=i \\ \alpha^{(i)} \prod_{l=g+1}^{i} \iota^{(l)}, & g<i\end{cases},
\end{align}
\noindent $p_{(g)}^{(i)}\triangleq -\vartheta_{(g)}^{(i)}w_{i-g}$, and $\varepsilon^{(i)}\triangleq -\sum_{g=1}^{i}p_{(g)}^{(i)}$. Therefore, $c_{0}^{(i)}=-\sum_{g=1}^{i-1}p_{(g)}^{(i)}/w_{0}$, $c_{1}^{(i)}=\sigma^2 w_{0}/\rho L+(v_{m}^{(i)})^{2} \bar{w}_{0,0}$, $c_{2}^{(i)}=-\sum_{g=1}^{i-1} \vartheta_{(g)}^{(i)}\left(\sigma^2 w_{i-g}/\rho L+\operatorname{Re}\left((v_{m}^{(i,g)})^{2}\right) \bar{w}_{0, i-g}\right)$ with $\operatorname{Re}\left((v_{m}^{(i,g)})^{2}\right)$ denoting the real part of the covariance $\frac{\mathbb{E}[(\mathbf{s}_m^{(i)}-\mathbf{h}_m)^{\mathrm{H}}(\mathbf{s}_m^{(g)}-\mathbf{h}_m)]}{(T+1) N}$, and $c_{3}^{(i)}=\sum_{g=1}^{i-1} \sum_{l=1}^{i-1} \vartheta_{(g)}^{(i)} \vartheta_{(l)}^{(i)}\left(\sigma^2 w_{2 i-g-l}+(v_{m}^{(g,l)})^{2} \bar{w}_{i-g, i-l}\right)$.

It can be observed from (\ref{MAMPLEB}) that the matched filter $\mathbf{P}^{\mathrm{H}}$ plays a similar role as the LMMSE estimator in the OAMP-based algorithm. To guarantee the orthogonality between the input and output estimation errors of the LE and NLE \cite{leiliu2022}, all messages obtained in preceding iterations, i.e., $\sum_{g=1}^{i}p_{(g)}^{(i)}\mathbf{s}_{m}^{(g)}$, are utilized. With the application of such a memory mechanism, matrix multiplications instead of matrix inversions are required in each MAMP iteration, which is the engine that substantially reduces the computational complexity. It can be shown that the computational complexity of each MAMP iteration is $\mathcal{O}(L(T+1)N)$, which is much lower than that of an OAMP iteration given as $\mathcal{O}(L^{2}(T+1)N)$. 

\section{Simulation Results}
We simulate a single-cell uplink cellular network with 400 users uniformly distributed within a circular ring. The path loss of user $n$ is modeled as $\beta_{n}=-128.1-36.7\log_{10}(d_{n})$ (dB) with $d_{n} \in [0.05, 1]$ km. The number of BS antennas is $M=16$, the pilot sequence length is $\bar{L}=50$, and the maximum delay is $T=4$ symbols. Besides, the transmit power of each user is set to be $23$ dBm, and the noise power spectrum density is $-169$ dBm/Hz over $1$ MHz bandwidth. In addition, the maxmium number of iterations is $Q_{1}=50$, the accuracy tolerance is $\epsilon_{1}=10^{-5}$, and the empirical threshold for activity detection is $\theta=0.6$. The simulation results are averaged over $10^5$ independent channel realizations. For comparisons, the following two baselines are also simulated:
\begin{itemize}
    \item \textbf{Group LASSO-based method \cite{lliu2021}:} This scheme formulates the joint activity-delay detection and channel estimation as a group LASSO problem, which is solved by a block coordinate descent algorithm. 
\end{itemize}
\begin{itemize}
    \item \textbf{AMP-based method \cite{wzhu2021}:} This method uses the AMP algorithm with an MMSE denoiser to perform joint activity-delay detection and channel estimation. However, the deep learning implementation in \cite{wzhu2021} is not considered in this paper due to the marginal gain of detection and estimation accuracy.
\end{itemize}

\begin{figure}[t]
\centering
\includegraphics[width=2.5in]{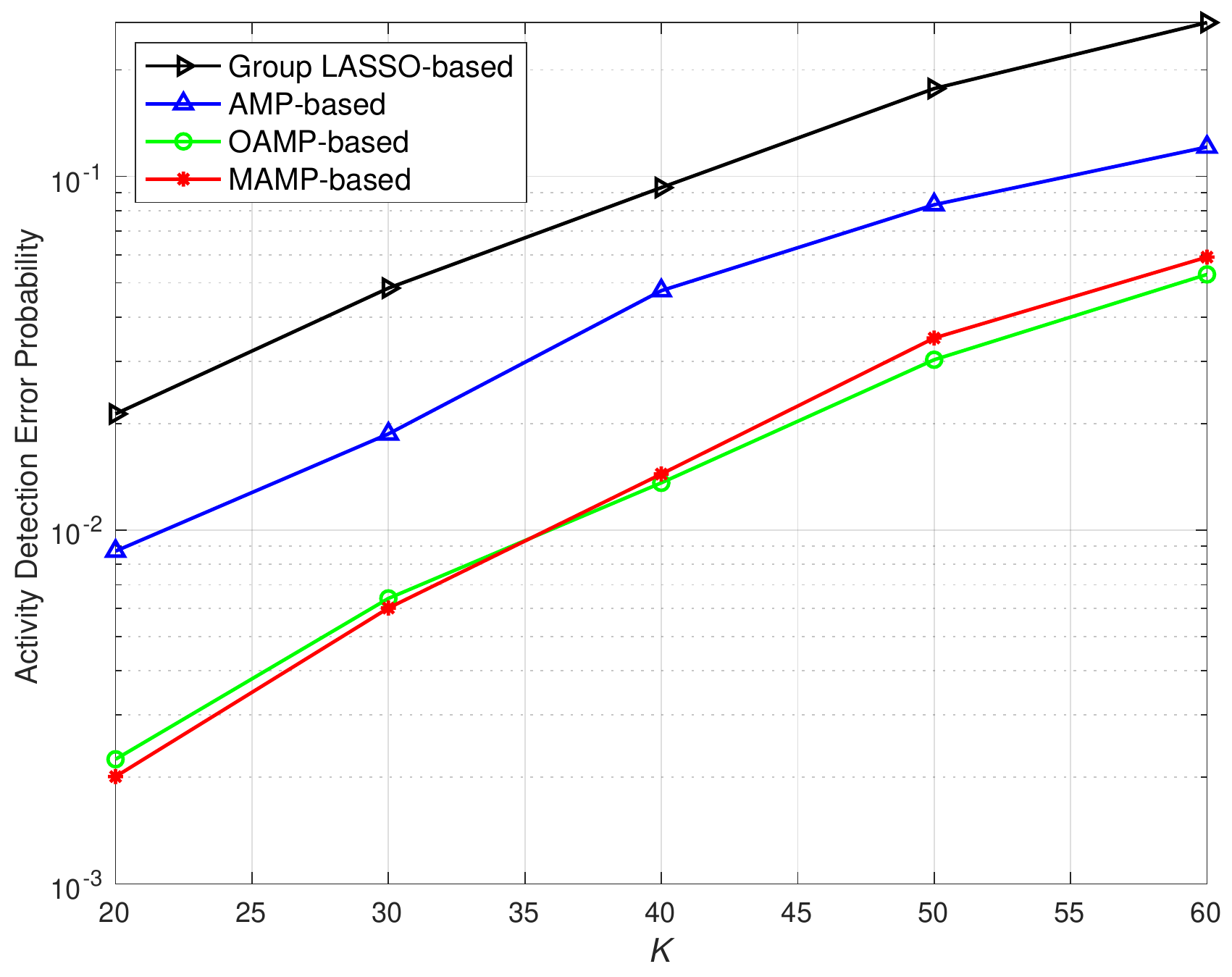}
\caption{Activity detection error probability versus the number of active users.}
\label{audvk}
\end{figure}

\begin{figure}[t]
\centering
\includegraphics[width=2.5in]{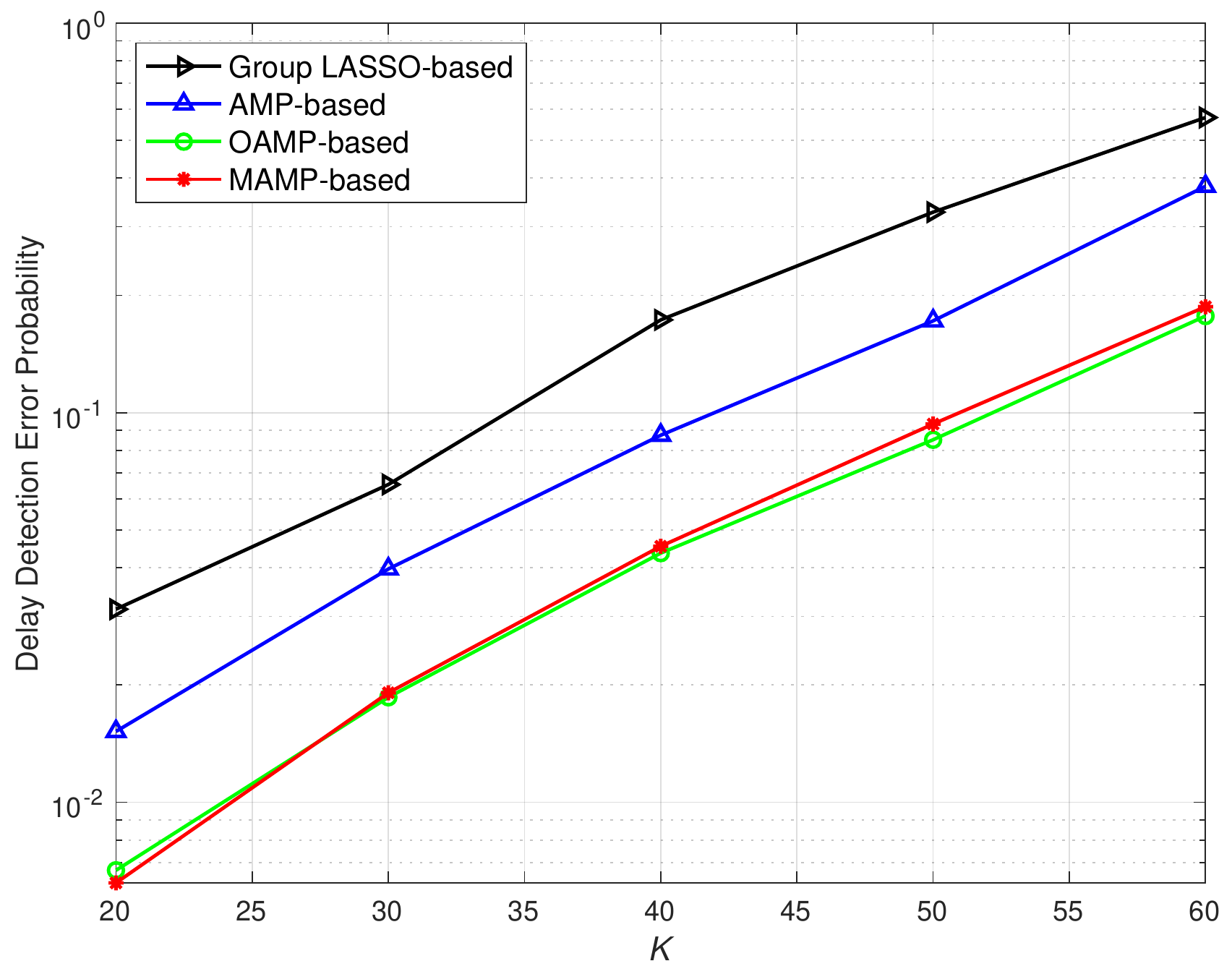}
\caption{Delay detection error probability versus the number of active users.}
\label{delayvk}
\end{figure}

We first show the activity detection and delay detection error probability versus the number of active users in Fig. \ref{audvk} and Fig. \ref{delayvk}, respectively. It is observed from both figures that the detection accuracy degrades with the number of active users due to the limited radio resources for pilot transmissions. Compared to the baseline methods, the two proposed algorithms effectively reduces the user activity detection and delay detection error since both the OAMP- and MAMP-based algorithm are capable of handling pilot matrices with non-i.i.d. entries. The performance gain is also attributed to the exploration of the common sparsity pattern among multiple BS antennas. Besides, performance of the proposed MAMP-based algorithm is practically the same as that achieved by the OAMP-based algorithm, which indicates the reduced complexity does not compromise its effectiveness.

\begin{figure}[t]
\centering
\includegraphics[width=2.5in]{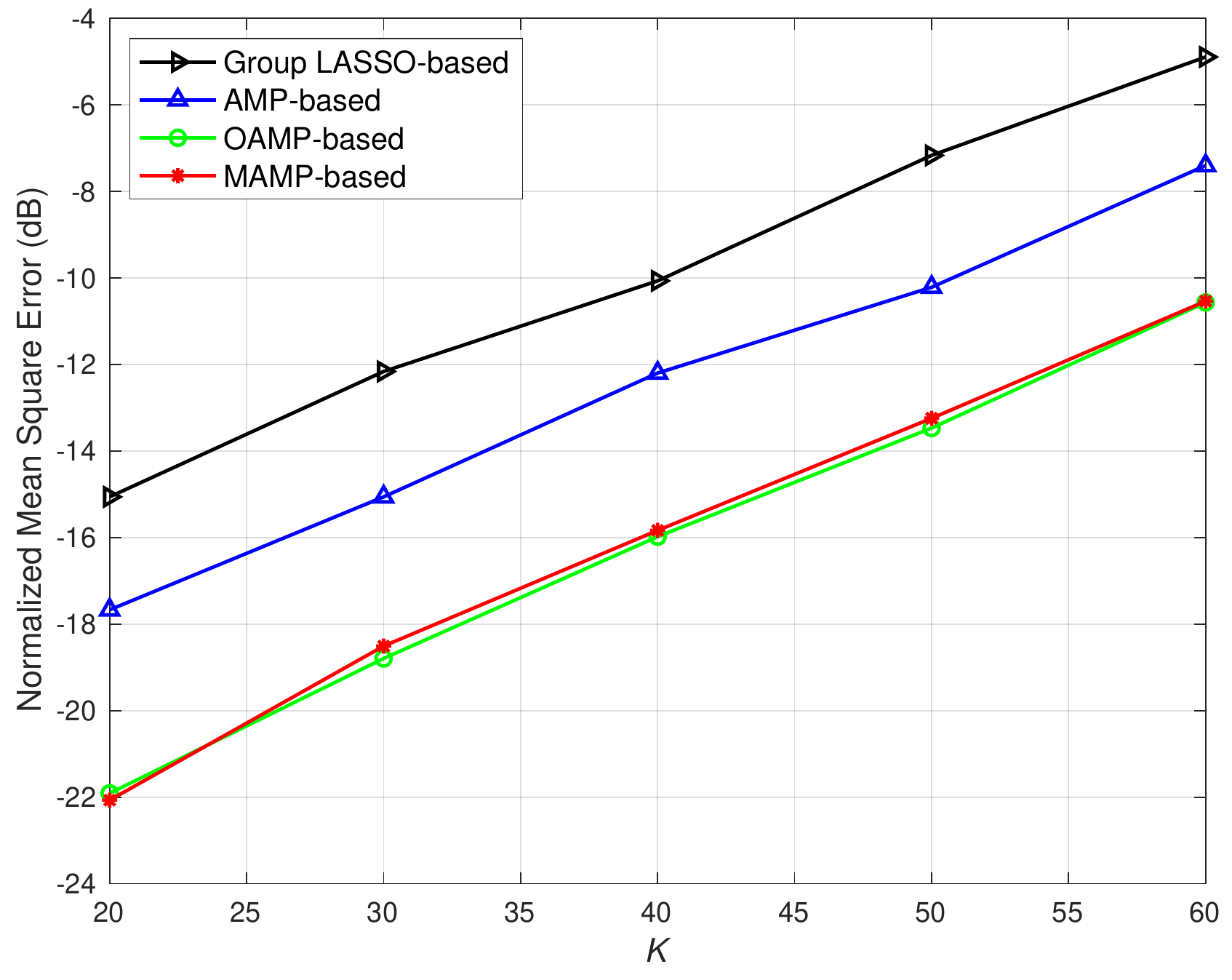}
\caption{NMSE of channel estimation versus the number of active users.}
\label{nmse}
\end{figure}

Next, we investigate the relationship between the normalized mean square error (NMSE) of channel estimation and the number of active users in Fig. \ref{nmse}. Similar to Figs. \ref{audvk} and \ref{delayvk}, the two proposed algorithms achieve significant NMSE reduction compared with the two baselines, while the MAMP-based algorithm maintains similar performance as the OAMP-based algorithm. These observations again validate the benefits of the LMMSE estimator of OAMP, and the memory LE as well as the divergence-free NLE in MAMP, which make the AMP-type algorithms compatible with more general pilot matrices.

We further examine the computation complexity of different algorithms by measuring their average execution time on the same computing server. The number of active users is fixed as $K=50$, and the results are summarized in Table \ref{time}. From the table, it is clear that the group LASSO-based method has the lowest complexity. However, its performance is far worse than other methods. Besides, while the OAMP-based algorithm achieves the best detection and estimation accuracy, it suffers from heavy computational overhead that originates from the LMMSE estimator. In addition, the AMP- and MAMP-based algorithm have comparable average execution time, and the MAMP-based algorithm secures a 37\% complexity reduction compared with the OAMP-based method. This demonstrates the superiority of the memory mechanism in the MAMP-based algorithm in achieving low complexity and satisfactory detection/estimation accuracy.

\section{Conclusions}
In this paper, we investigated the joint activity detection, synchronization delay detection, and channel estimation in asynchronous grant-free massive random access systems. Considering that entries in the pilot matrix are not independent and identically Gaussian distributed, we first proposed a novel algorithm based on the orthogonal approximate message passing (OAMP), which also makes full utilization of the common sparsity among the received pilot signals of multiple base station antennas. To accelerate the computation, a memory approximate message passing (MAMP)-based algorithm was further developed, which introduces a memory mechanism to avoid matrix inversion. Simulation results showed the effectiveness of the proposed algorithms, and the potential of the MAMP-based algorithm for fast joint activity-delay detection and channel estimation in grant-free massive random access (RA) systems. Our study also advocates the developments of more advanced algorithms for asynchronous massive RA systems, e.g., by fusing deep unrolling and MAMP, to further enhance the performance and reduce the complexity.
\begin{table}[t]
\caption{Average execution time of different methods}
\centering
\begin{tabular}{c|c}
    \hline
     Methods & Average execution time (s)  \\
    \hline
    Group LASSO-based & 3.82  \\
    \hline
    AMP-based & 4.16 \\
    \hline
    OAMP-based & 6.94\\
    \hline
    MAMP-based & 4.31\\
    \hline
    \end{tabular}
\label{time}
\end{table}


\begin{thebibliography}{00}
\bibitem{cbo2016} C. Bockelmann \emph{et al.}, “Massive machine-type communications in 5G: Physical and MAC-layer solutions,” \emph{IEEE Commun. Mag.}, vol. 54, no. 9, pp. 59–65, Sep. 2016.
\bibitem{mt2014} M. T. Islam, A. E. M. Taha, and S. Akl, “A survey of access management techniques in machine type communications,” \emph{IEEE Commun. Mag.}, vol. 52, no. 4, pp. 74–81, Apr. 2014.
\bibitem{ywu2020} Y. Wu \emph{et al.}, “Massive access for future wireless communications," \emph{IEEE Wireless Commun.}, vol. 27, no. 4, pp. 148-156, Aug. 2020.
\bibitem{psch2017} P. Schulz \emph{et al.}, “Latency critical IoT applications in 5G: Perspective onthe design of radio interface and network architecture,” \emph{IEEE Commun. Mag.}, vol. 55, no. 2, pp. 70-78, Feb. 2017.
\bibitem{xchen2021} X. Chen, D. W. K. Ng, W. Yu, E. G. Larsson, N. Al-Dhahir, and R. Schober, “Massive access for 5G and beyond,” \emph{IEEE J. Sel. Areas Commun.}, vol. 39, no. 3, pp. 615–637, Mar. 2021.
\bibitem{donoho2006} D. L. Donoho, “Compressed sensing,” \emph{IEEE Trans. Inf. Theory}, vol. 52, no. 4, pp. 1289–1306, Apr. 2006.
\bibitem{trob1996} T. Robert, “Regression shrinkage and selection via the Lasso,” \emph{J. Roy. Statist. Soc.}, vol. 58, no. 1, pp. 267–288, Jan. 1996.
\bibitem{jat2004} J. Tropp, “Greed is good: Algorithmic results for sparse approximation,” \emph{IEEE Trans. Inf. Theory}, vol. 50, no. 10, pp. 2231–2242, Oct. 2004.
\bibitem{donoho2010} D. L. Donoho, A. Maleki, and A. Montanari, “Message passing algorithms for compressed sensing: I. motivation and construction,” in \emph{Proc. IEEE Inf. Theory Wkshop. (ITW)}, Cairo, Egypt, Jan. 2010.
\bibitem{yafeng2023} Y.-F. Liu, W. Yu, Z. Wang, Z. Chen, and F. Sohrabi, “Grant-free random access via covariance-based approach,” [Online]. https://www.comm.utoronto.ca/weiyu/2023\underline{\ }Chapter\underline{\ }Covariance.pdf
\bibitem{sha2018} S. Haghighatshoar, P. Jung, and G. Caire, “Improved scaling law for activity detection in massive MIMO systems,” in \emph{Proc. IEEE Int. Symp. Inf. Theory (ISIT)}, Vail, CO, USA, Jun. 2018.
\bibitem{zchen2019} Z. Chen, F. Sohrabi, Y.-F. Liu, and W. Yu, “Covariance based joint activity and data detection for massive access with massive MIMO,” in \emph{Proc. IEEE Int. Conf. Commun. (ICC)}, Shanghai, China, May 2019.
\bibitem{zchen2018} Z. Chen, F. Sohrabi, and W. Yu, “Sparse activity detection for massive connectivity,” \emph{IEEE Trans. Signal Process.}, vol. 66, no. 7, pp. 1890–1904, Apr. 2018.
\bibitem{mke2020} M. Ke, Z. Gao, Y. Wu, X. Gao, and R. Schober, “Compressive sensing based adaptive active user detection and channel estimation: Massive access meets massive MIMO," \emph{IEEE Trans. Signal Process.}, vol. 68, pp. 764–779, 2020.
\bibitem{xbian2022} X. Bian, Y. Mao, and J. Zhang, “Joint activity detection, channel estimation, and data decoding for grant-free massive random access,” \emph{IEEE Internet Things J.}, to appear.
\bibitem{hongyuan2008} H. Zhang \emph{et al.}, “Asynchronous interference mitigation in cooperative base station systems,” \emph{IEEE Wireless Commun.}, vol. 7, no. 1, pp. 155-165, Jan. 2008.
\bibitem{lliu2021} L. Liu, and Y. Liu, “An efficient algorithm for device detection and channel estimation in asynchronous IoT systems,” in \emph{Proc. IEEE Int. Conf. Acoustics, Speech Signal Process. (ICASSP)}, Jun. 2021.
\bibitem{wzhu2021} W. Zhu, M. Tao, X. Yuan, and Y. Guan, “Deep-learned approximate message passing for asynchronous massive connectivity,” \emph{IEEE Trans. Wireless Commun.}, vol. 20, no. 8, pp. 5434-5448, Mar. 2021.
\bibitem{jma2017} J. Ma and L. Ping, “Orthogonal AMP,” \emph{IEEE Access}, vol. 5, pp. 2020–2033, 2017.
\bibitem{ymei2022} Y. Mei \emph{et al.}, “Compressive sensing-based joint activity and data detection for grant-free massive IoT access,” \emph{IEEE Trans. Wireless Commun.}, vol. 21, no. 3, pp. 1851-1869, Mar. 2022.
\bibitem{leiliu2022} L. Liu, S. Huang, and B. M. Kurkoski, “Memory AMP,” \emph{IEEE Trans. Inf. Theory}, vol. 68, no. 12, pp. 8015-8039, Dec. 2022.
\bibitem{ychi2019} L. Liu, Y. Chi, C. Yuen, Y. L. Guan, and Y. Li, “Capacity-achieving MIMO-NOMA: Iterative LMMSE detection,” \emph{IEEE Trans. Signal Process.}, vol. 67, no. 7, pp. 1758–1773, Apr. 2019.
\end{thebibliography}
\end{document}